\documentclass{sadhana}

\usepackage{graphicx,amsmath,bm}
\usepackage{multirow}

\newcommand{\etal}{\textit{et al.}}

\begin{document}

\title{Initial Decoding with Minimally Augmented Language Model for Improved Lattice Rescoring in Low Resource ASR}


\author{Savitha Murthy\textsuperscript{1,*}\and Dinkar Sitaram\textsuperscript{2}}
\affilOne{\textsuperscript{1} Department of CSE, P.E.S. University\\}
\affilTwo{\textsuperscript{2} Cloud Computing Innovation Council of India}

\twocolumn[{

\maketitle

\begin{abstract}
Automatic speech recognition systems for low-resource languages typically have smaller corpora on which the language model is trained. Decoding with such a language model leads to a high word error rate due to the large number of out-of-vocabulary words in the test data. Larger language models can be used to rescore the lattices generated from initial decoding. This approach, however, gives only a marginal improvement. Decoding with a larger augmented language model, though helpful, is memory intensive and not feasible for low resource system setup. The objective of our research is to perform initial decoding with a minimally augmented language model. The lattices thus generated are then rescored with a larger language model. We thus obtain a significant reduction in error for low-resource Indic languages, namely, Kannada and Telugu. 

This paper addresses the problem of improving speech recognition accuracy with lattice rescoring in low-resource languages where the baseline language model is not sufficient for generating inclusive lattices. We minimally augment the baseline language model with unigram counts of words that are present in a larger text corpus of the target language but absent in the baseline. The lattices generated after decoding with a minimally augmented baseline language model are more comprehensive for rescoring. We obtain 21.8\% (for Telugu) and 41.8\% (for Kannada) relative word error reduction with our proposed method. This reduction in word error rate is comparable to 21.5\% (for Telugu) and 45.9\% (for Kannada) relative word error reduction obtained by decoding with full Wikipedia text augmented language mode while our approach consumes only 1/8th the memory. We demonstrate that our method is comparable with various text selection-based language model augmentation and also consistent for data sets of different sizes. Our approach is applicable for training speech recognition systems under low resource conditions where speech data and compute resources are insufficient, while there is a large text corpus that is available in the target language. Our research involves addressing the issue of out-of-vocabulary words of the baseline in general and does not focus on resolving the absence of named entities. Our proposed method is simple and yet computationally less expensive.
\end{abstract}


\keywords{Indic languages, Telugu and Kannada ASR, Low resource, out of vocabulary, language model augmentation, Automatic Speech Recognition}

}]


\corres \footnote{savithamurthy@pes.edu}
\articleType{}


\markboth{}{Minimal LM Augmentation for Effective Rescoring}

\begin{sloppypar}
\section{Introduction}
There is a lot of interest in Automatic Speech Recognition (ASR) for low-resource languages for more than a decade \cite{hazen2009query, novotney2010cheap, thomas2010cross}. Low resource indicates a scarcity of any of the language resources required to train traditional ASR, namely, resources such as pronunciation dictionaries, text data to train a Language Model (LM) or audio data with corresponding transcriptions. Obtaining low Word Error Rates (WER) in low-resource ASR is a challenge. These ASR systems have low to medium vocabulary of only a thousand to less than 50,000 words. This results in a high probability of Out-Of-Vocabulary (OOV) words that may be present in the test set leading to high WER. Data augmentation has been a predominant approach in improving ASR performance for low-resource languages. The prevalence of enormous amounts of data on the World Wide Web has made it possible to leverage more data for data augmentation purposes \cite{bulyko2007web, mendels2015improving, sethy2006text, parada2010spoken, ng2005web}.

Our study involves augmenting the language model using larger text corpora from the web in a traditional hybrid ASR setup for low-resource Indic languages, namely, Telugu and Kannada. A common approach in traditional ASR systems is to train a transcript language model from the available speech transcripts of the training dataset. In the case of low-resource languages, language models trained on speech corpus only with a few hours of training data may contain many unknown or OOV words. Decoding with such small language models leads to high WER. This issue can be addressed by interpolating the baseline language model with a larger language model in the concerned language to account for as many words as possible. A larger language model though comprehensive requires the construction of a decoding graph in a traditional hybrid ASR setup and is very memory intensive. In case of limited availability of computational resources, allocating the required memory may not be feasible. Alternatively, the availability of a larger text corpora can be harnessed by performing initial decoding with a smaller language model to generate a lattice and then re-scoring the lattice with the larger well-trained language model \cite{beck2019lstm}. However, in the case of low-resource languages, a small language model trained on the available speech text may not be sufficient to generate a comprehensive lattice. This leads to a large number of missing words in the lattices which are not resolved even after rescoring using a large, augmented language model.

\subsection{Our Contribution}
In this paper, we focus on improving OOV recovery on the baseline and thus making lattice rescoring with a larger language model more effective for low-resource languages. We consider the baseline language model augmented with Wikipedia text for our experiments using a larger language model. We explore language model augmentation techniques for two Indian languages namely, (i) Kannada — a seed corpus with 4 hours of read speech and (ii) Telugu — 40 hours speech corpus of read and conversational speech. We train the baseline language model on the available speech transcripts. We train another language model on the unigram counts of the Out-of-Train (OOT) words from the larger text corpus that are not present in the baseline train transcript. We augment the baseline language model by interpolating it with the word unigram language model. We perform initial decoding with this minimally augmented language model. We separately train a language model using Wikipedia text and interpolate it with the baseline to obtain a larger language model. The lattices generated from initial decoding are then rescored with the larger language model. Our experiments include different text selection methods as well as different-sized datasets. Our method also eliminates the empirical need to determine the amount of text that needs to be selected for language model augmentation purposes.

Our empirical study regarding language model augmentation and lattice rescoring in low-resource languages depicts the following:
\begin{enumerate}
\item Lattices generated after decoding with the baseline language model may not contain all the probable words (earlier OOVs). Lattice rescoring only adjusts the probabilities on the existing path and hence there is no significant improvement in OOV recovery as well as WER.
\item \label{item2} Initial decoding with a baseline language model augmented to include unigram counts of OOT words (words that are present in the bigger text corpus but not in the baseline vocabulary) improves OOV recovery and hence enables the inclusion of more words in the lattices.
\item Rescoring the lattices generated from \ref{item2} with Wikipedia augmented language model results in a significant reduction in WER. This reduction is comparable to that of decoding with a larger language model while re-scoring is more time and memory efficient.
\item In-Vocabulary (IV) recognition is not affected and only improves using our proposed method.
\end{enumerate}

This paper is organized as follows: section \ref{sect:related_work} gives an overview of different data augmentation techniques used in ASR – for both Large Vocabulary Speech Recognition (LVSCR) and low resource languages. It also summarizes the OOV recovery techniques used. Section \ref{sect:problem} describes the OOV problem in the context of low-resource languages namely, Telugu and Kannada. Section \ref{sect:concept} explains the concept of decoding and lattice rescoring in brief. Section \ref{sect:data_setup} describes the datasets used for our experiments and the experimental setup. Section \ref{sect:exp} gives an overview of our experiments. Section \ref{sect:result} discusses the results followed by a section \ref{sect:conclusion} on the conclusion and future work.

\section{Related Work}
\label{sect:related_work}
This section gives an overview of various data augmentation techniques that have been employed for improving recognition accuracy in ASR in section \ref{sect:augmentation} and an overview of OOV detection and recovery literature in section \ref{sect:oov}.

\subsection{Data Augmentation in ASR}
\label{sect:augmentation}  
Data augmentation is one of the research directions towards improving the performance of ASR systems. Various approaches of data augmentation include (i) Perturbation (ii) Use of mixed features (iii) Transfer learning (iv) Text augmentation (v) Leveraging multilingual features (vi) Speech synthesis/generation. Perturbation is a technique where the training data is augmented with variations such as swapping blocks of time steps and frequencies~\cite{song2020specswap}, changing the speed of audio signal~\cite{ko2015audio, vachhani2018data} and modifying vocal tract length~\cite{tuske2014data, jaitly2013vocal}. Chen \etal~\cite{chen2020data} employ different perturbations such as pitch, speed, tempo, volume, reverberation and spectral augmentation to improve the accuracy of children's speech recognition. Mix-up is another technique for data augmentation wherein more samples of training data are generated using a convex combination of pair of samples and their labels~\cite{medennikov2018investigation, saon2019sequence, zhu2019mixup}. Transfer learning is a technique where model parameters are shared across multiple tasks and can be considered as a method for data augmentation~\cite{ghahremani2017investigation, manohar2017jhu, wang2020cross}. Speech synthesis/generation has recently been used to increase the amount of training data~\cite{li2018training}. The text-to-speech system is trained on ASR training data to match the style of the training corpus for better recognition \cite{laptev2020you}. Rosenberg \etal~\cite{rosenberg2019speech} also train a Tacotron model on Librispeech corpus and out-of-domain speech to explore the effect of acoustic diversity. Text augmentation has been used \cite{benevs2020text, peyser2020improving} to improve language model scores and reduce WER. The aforementioned research has been adapted for languages with sufficient resources. 

A few of these aforementioned augmentation techniques have also been applied to low-resource ASR to improve accuracy. Sharma \etal~\cite{sharma2020improving} use mixup technique to augment speech data with TTS audio while Meng \etal~\cite{meng2021mixspeech} use mixed features (MFCC and mel-spectrograms) for data augmentation to train Listen, Attend and Spell (LAS) and Transformer models. Rossenbach \etal~\cite{rossenbach2020generating} improve the performance of an End-to-End (E2E) ASR by augmenting the acoustic model with synthesized speech generated using a Text To Speech (TTS) system trained on the ASR speech corpus. The work by Lin \etal~\cite{lin2020training} employs synthesized speech to improve keyword spotting with limited data. Perturbation has been used to augment features for speech training~\cite{hartmann2016two, hailu2020improving, rebai2017improving}. A multilingual approach where speech data from similar languages are leveraged to augment and train the acoustic model is quite popular in low-resource ASR. Rosenberg \etal~\cite{rosenberg2017end} applied a multilingual approach to improving keyword search in IARPA BABEL OP3 languages. The `Low Resource Speech Recognition Challenge for Indian Languages - Interspeech 2018' included 40 hours of speech corpora in Telugu, Tamil and Gujarati languages. Multilingual training was adapted wherein the acoustic model was trained in all three languages leading to an improvement of approximately 5-8\% in WER~\cite{vydana2018exploration, fathima2018tdnn, pulugundla2018but, shetty2018articulatory, billa2018isi}. However, these methods above reduce recognition errors in words already present in the ASR's lexicon. A combination of methods, namely, vocal tract length perturbation and multilingual features are adapted for IARPA Babel languages in the work by Tuske \etal~\cite{tuske2014data}. Y{\i}lmaz \etal~\cite{yilmaz2018acoustic} improve ASR in code-switched speech for Dutch and Frisian. They augment the language model by generating text using Recurrent Neural Network (RNN) LM trained on the transcripts from the code-switched speech data. They also enhance the audio from available high-resource data in code-switched speech.

\subsection{OOV Detection and Recovery}
\label{sect:oov}
There has been extensive research on OOV detection and recovery to improve ASR recognition accuracy. The challenge in handling OOVs is that the ASR system is unaware of the presence of an OOV. This, in turn, results in the hypothesis always containing words in the lexicon leading to errors in recognition. There have been efforts to detect OOVs using filler models~\cite{klakow1999oov, schaaf2001detection, kitaoka2021dynamic} where there are placeholders for OOVs which can then be replaced with the extended vocabulary for improved recognition. Confidence measures are another indication of OOV presence. Confidence scores are used to identify the probable OOV candidates~\cite{thomas2019detection, hazen2001comparison, yazgan2004hybrid, ketabdar2007detection, white2008confidence, rastrow2009new}. Studies have tried to achieve open vocabulary ASR by employing subword models where hybrid language models - both word LM and subsequence (phoneme, syllable or subword) are used together. In this case, the unknown words are replaced with subsequences to aid better recognition. The most recent work is by Zhang~\etal~\cite{zhang2020oov} where a hybrid lexical model with phonemes and words is used to generate OOV candidates with phoneme constraints. Instead of hybrid subword models, parallel models also have been used for OOV detection and recovery where both subsequence lattices and word lattices are used to determine OOV~\cite{hazen2001comparison, lin2007oov, burget2008combination, kombrink2009posterior, murthy2018effect}. OOV recovery follows OOV detection. P2G mapping is the most popular for OOV recovery where the subsequences are stitched together to form words using P2G models~\cite{hori2017multi, zhang2020oov}. Another approach is clustering or similarity-based approach where words similar to reference or in-vocabulary words to determine OOV probabilities~\cite{hannemann2010similarity, inbook, DBLP:conf/icassp/EgorovaB18, DBLP:conf/interspeech/QinSR11, DBLP:journals/ieicet/NaptaliTN12}.

While data augmentation has been predominantly used to reduce WER in LVSCR, very few researchers adapt data augmentation to handle OOV in ASR~\cite{DBLP:conf/icnlsp/OrosanuJ15, DBLP:conf/icassp/WangZLWDQ21, DBLP:conf/icassp/ZhengLGW21}. These studies address issues related to specific words such as proper nouns. Others select sentences using distance measures to ensure the proximity of selected sentences to the training data. Naptali \etal~\cite{DBLP:journals/ieicet/NaptaliTN12} use web data to determine the OOV probabilities after determining OOV candidates based on similarity measures. These approaches are effective when there is a sufficient amount of training data.

These approaches mentioned above to handle OOV detection and recovery, may not be very effective for languages that are agglutinative and inflective, with every root word having several forms based on different contexts. It is all the more challenging to address the issue of OOV detection and recovery when the language in concern is low resource and lacks sufficient data to train an ASR. We explain the problem of OOV in such languages in the following section.

\section{OOV Problem in Low Resource Agglutinative and Inflective Languages}
\label{sect:problem}
Low-resource languages have high OOV rates due to limited vocabulary and language model size. This is more prominent in the case of agglutinative languages. Dravidian languages (languages spoken in southern India) like Telugu and Kannada are agglutinative, resulting in words of different lengths derived from the basic word forms. They are also highly inflective for the following reasons: (a)There are 8 cases (called \emph{vibhaktis}) in Telugu with 3 to 4 suffixes corresponding to each of the cases resulting in a total of 30 suffixes as per the \emph{vibhaktis} of each noun form in Telugu; (b)There are different suffixes for two genders (called \emph{linga}) in Telugu which again differ based on the number which is singular or plural; (c)Telugu being highly inflective the nouns change forms based on the case, gender and number resulting in 30 * 3 * 2 = 150 variations for each noun in the language; (d)There are 10 different \emph{sandhis} in Telugu, where multiple words are joined together to form a single word along, with 7 \emph{sandhis} that come from the Sanskrit language; (e) The suffix for verb (all tenses) in a sentence is based on the gender, number and case of the noun it corresponds to resulting in higher inflection. Kannada also has similar inflections with eight cases and three genders, with noun forms being inflected based on cases, gender, and tense. Also, \emph{sandhis} in Kannada (3 in Kannada + 7 borrowed from Sanskrit) again can join multiple words to form single words.

The problem of OOV for such languages does not only include addressing named entities which are always of concern in all the languages, but also the need for a large corpus with a comprehensive vocabulary that accounts for all the agglutinative and inflective forms of nouns, verbs, adjectives and so on. For example, the word `ABHIMAANISTUNNAANANI' \footnote{The example words in Telugu are represented using IITM label set notation \cite{shetty2021exploring}}  in Telugu that occurs in the test data is not present in Wikipedia. The root form of this word is `ABHIMAANISTUNNA', and a different inflection `ABHIMAANISTUNNAARU' of the word is present. Defining a text corpus that includes all inflections of a word requires linguistic expertise, and obtaining corresponding audio is a tedious task. Hence, speech corpora for these languages tend to have high OOV rates. There has been work on improving ASR recognition accuracy in agglutinative languages by splitting the words into morphemes. The focus of our experiments is to address OOV recovery and WER reduction for any low-resource language in general and, is complementary to the technique of using morphemes as subwords \cite{creutz2007morph, lileikyte2018conversational, he2014subword, choueiter2006morpheme}.

\section{Concept of Decoding and Lattice Rescoring}
\label{sect:concept}
\subsection{Decoding}
Decoding involves finding the most probable path of word sequences. A Weighted Finite State Transducer (WFST)~\cite{povey2012generating} is used for this purpose. It is represented as given in Equation\ref{decoding}:
\begin{equation}
\label{decoding}
HCLG = min(det(H \circ C \circ L \circ G))
\end{equation}
where H represents the HMM transitions, C represents context dependencies among phones, L represents the lexicon and G represents the grammar represented by the language model.

While decoding with a bigger language model may be more effective in reducing WER, composing an HCLG graph, referred to as a decoding graph, with the grammar belonging to a large language model is more memory intensive, and decoding becomes computationally expensive because of a large search space. Hence, normal practice is to decode using the smaller language model and then perform a lattice rescoring with a bigger language model.

\subsection{Lattice Rescoring}
Lattice represents alternate possible word sequences having higher scores than other possible word sequences. A lattice contains the paths with higher scores and is derived from the pruned subset of the decoding graph~\cite{povey2012generating}. Lattices are generated as a result of decoding. Lattice rescoring is a process where the path probabilities are updated with the new language model probabilities while retaining the transition and pronunciation probabilities. Lattice rescoring can be performed using a larger language model. This also conserves computational resources as there is no need to compose a decoding graph.

\section{Datasets and Experimental Setup}
\label{sect:data_setup}
\subsection{Datasets}
We consider (i) Telugu speech corpus released by Microsoft as part of the ``Speech Recognition for Indian Languages Challenge'' by Srivastava \etal at Interspeech, 2018 \cite{srivastava2018interspeech} and (ii) Kannada speech corpus recorded using the transcripts of speech synthesis dataset from IIIT Hyderabad\footnote{Accessed July 13, 2017. http://festvox.org/databases/iiit voices}  for our experiments. Telugu speech corpus consists of 40 hours of read and conversational speech. Every audio recorded in both datasets is 16 kHz mono. The Telugu training dataset has a vocabulary of 43,260 words, and the test dataset has an OOV rate of 12.04\%. The high OOV rate in Telugu is due to the reasons listed in section 3. Addressing these challenges requires a notably large dataset. Hence, the Telugu dataset used can be considered a low resource. The Kannada baseline vocabulary is 1754 words, and the test data has an OOV rate of 25.22\%. Kannada dataset with 4 hours of speech can be considered an extremely low resource corpus due to the reasons listed in section \ref{sect:problem}. Table \ref{tab:baseline} lists the baseline WER of Telugu (which is 25.51\%) and Kannada (which is 51.87\%) ASR systems.

\begin{table}[h!]
\caption{Telugu and Kannada ASR baselines}
\label{tab:baseline}
\begin{tabular}{|l|l|c|c|}
\hline
\bf{Language} & \bf{Duration}  & \bf{WER (\%)} & \bf{OOV rate (\%)}\\
\hline
Telugu & 40 hours & 25.51 & 12.04\\
Kannada & 4 hours & 51.87 & 31.58\\
\hline
\end{tabular}
\tablenotes{$-$ To avoid recognition errors due to agglutination, we concatenate consecutive words to form the longest word in the vocabulary.}
\tablenotes{$-$ WER for Telugu as specified in Microsoft release is 34.36\%. We obtained a WER of 25.51\% after test transcript corrections and accounting for agglutination.}
\label{tab:baseline}
\end{table}

\subsection{Experimental Setup}
The baseline ASR consists of TDNN-F~\cite{povey2018semi} acoustic model with 128 chunks, 7 hidden layers, 2048 hidden dimensions, an initial learning rate of 0.008 and 10 epochs using Kaldi ASR toolkit~\cite{povey2011kaldi} recipe.

The language model for the baseline is trained on the speech transcripts in the training set. Trigram language model with Witten bell smoothing is employed. We use the `count merging' method~\cite{pusateri2019connecting}\cite{hsu2007generalized} of interpolation to augment the baseline language model. Count merging is given by Equation \ref{count_merge}
\begin{equation}
\label{count_merge}
p^{CM}(w|h) = \frac{\sum_i\beta_ic_i(h)p_i(w|h)}{\sum\beta_jc_j(h)}
\end{equation}
where $i$ represents the domain or the model $\beta_i$ is the model scaling factor for $i$ and $h$ is the history for the $i^{th}$ model or domain. The interpolation weight for a given history in count merging is given as:
\begin{equation}
\lambda_i(h) = \frac{\beta_ic_i(h)}{\sum_j\beta_jc_j(h)}
\end{equation}

The interpolation weight in count merging depends on the history counts pertaining to a particular domain instead of absolute counts. Thus, the history counts that belong to a smaller corpus get more weightage as compared to the same history in a larger corpus. Count merging is stated to perform better than linear interpolation~\cite{pusateri2019connecting}. We compare HCLG graph decoding and baseline lattice rescoring methods with the augmented LM.

\section{Experiments}
\label{sect:exp}
This section describes various experiments performed concerning language model augmentation. Section \ref{sect:full_wiki} describes the details of augmenting the language model with full Wikipedia text for Telugu and Kannada languages. Section \ref{sect:memory} gives details of language model augmentation with a focus on OOT in terms of memory requirements and the reason for choosing only unigram counts of OOT words for minimally augmenting the baseline language model. Section \ref{sect:selection} describes various text selection techniques present in literature that we employ for augmentation purposes. Section \ref{sect:datasize} lists details on various dataset sizes used for our experiments.

\subsection{Language Model Augmentation with full Wikipedia text}
\label{sect:full_wiki}
We use Wikipedia text data in Telugu and Kannada to enhance the corresponding Language Models (LM). The text data was pre-processed to remove non-alphabet characters and was normalized. Telugu Wikipedia XML dump data consisting of 2525122 sentences, was used to augment the language model that increased the vocabulary to 1815924 words from 43260 words in the baseline. Kannada Wikipedia XML dump consists of 873339 sentences increasing the vocabulary to 943729 words from an initial vocabulary of 1754 in the baseline.

\subsection{OOT-based Language Model Augmentation}
\label{sect:memory}
The memory requirement for decoding graph construction with a full Wikipedia text augmented language model is approximately 32 gigabytes for Telugu and 18 gigabytes for Kannada, as listed in Table \ref{tab:memory}, which is quite large. Therefore, we consider only OOT-based enhancements to the baseline language model. OOT words, in this context, are those words that are present in Wikipedia but not in the baseline training text. Table \ref{tab:memory} lists the different OOT-based enhancements employed and their memory requirements.

First, we consider Wikipedia lines containing OOT words for enhancing the baseline language model. In low-resourced, agglutinative, and inflective languages like Telugu and Kannada, lines containing OOT words constitute more than 90\% of the Wikipedia text. Augmenting the baseline language model with such a large subset requires approximately the same memory as full Wikipedia augmentation.

We then consider trigrams from Wikipedia containing OOT words. Again because of high OOT rates, this resulted in a bigger language model than the complete Wikipedia itself. As specified in Table \ref{tab:memory}, `baseline + Wikipedia OOT trigrams' requires more memory than `baseline + full Wikipedia' for both Telugu and Kannada.

After this, we consider augmenting the baseline language model with unigram counts of only OOT words. As seen in Table \ref{tab:memory}, such an augmentation has fewer memory requirements, comparatively very close to the memory need of the baseline language model. Therefore, we propose initial decoding with a graph composed of an LM augmented with OOT unigrams probabilities from Wikipedia. The generated lattices are then rescored with a bigger augmented language model. We compare our method against different text selection-based augmentation (described in section \ref{sect:selection}) as well as different data sizes (described in section \ref{sect:datasize}).

\begin{table*}[h!]
\caption{Memory requirements for decoding graph construction}
\label{tab:memory}
\begin{tabular}{|l|l|c|}
\hline
Language & Language Model & Max Memory Required\\
\hline
\multirow{5}{*} {Telugu} & baseline & $\sim$2GB\\
&baseline $+$ full Wikipedia & $\sim$32GB\\
&baseline $+$ Wikipedia OOT lines & $<$32GB\\
&baseline $+$ Wikipedia OOT trigrams & $>$32GB\\
&baseline $+$ Wikipedia OOT words & $~$4GB\\
\hline
\multirow{5}{*} {Kannada} & baseline & $\sim$1GB\\
&baseline $+$ full Wikipedia & $\sim$18GB\\
&baseline $+$ Wikipedia OOT lines & $\sim$18GB\\
&baseline $+$ Wikipedia OOT trigrams & $\sim$26GB\\
&baseline $+$ Wikipedia OOT words & $<$2GB\\
\hline
\end{tabular}
\tablenotes{$-$ GB: Giga Bytes}
\end{table*}

\subsection{Text Selection Based Language Model Augmentation}
\label{sect:selection}
Along with the complete Wikipedia text augmentation mentioned in section \ref{sect:full_wiki}, we adapt different text selection methods available in the literature, namely, contrastive selection~\cite{chen2020improving}, change in likelihood~\cite{klakow2000selecting}, entropy-based selection~\cite{itoh2012n}. These selection methods attempt to benefit from the availability of large amounts of non-domain data to improve language model probabilities, eventually improving ASR performance. We select 50\% of the highest-ranked sentences from Wikipedia for every text selection method used for language model augmentation.

\subsubsection{Contrastive Selection}
Chen \etal~\cite{chen2020improving} augment the language model with a contrastive selection from a larger text corpus. They select sentences that are similar to the ASR training set. We instead select trigrams with higher probability given the interpolated LM (baseline and Wikipedia) compared to probability given the Wikipedia LM as given in Equation ~\ref{equ:contrastive}
\begin{equation}
\label{equ:contrastive}
sentence\_score = \frac{\Sigma{[\log P(t|D) - \log P(t|B)]}}{\#(t)}
\end{equation}
where, $p(t|D)$ denotes trigram probability with respected to $D$ train and wiki interpolated language model, and $p(t|B)$ denoted trigram probability with respect to $B$ language model trained on Wikipedia. The sum of the differences in trigram probabilities is normalized by the number of trigrams in a sentence. This method selects sentences containing trigrams that are similar to the training text.

\subsubsection{Delta Likelihood Based Selection}
Klakow~\cite{klakow2000selecting} uses the change in the log-likelihood when a sentence is removed from the corpus, and the language model is trained on it. The work claims an improvement in perplexity and OOV rate with this approach. We implement this technique for trigrams represented in Equation~\ref{equ:delta}
\begin{equation}
\label{equ:delta}
\Delta S_i = \sum_w N_{target}(u,v,w) \log \frac{P(w|uv)}{P_{A_i}(w|uv)}
\end{equation}
where $target$ represents the train(baseline) corpus, $u,v$ and $w$ are words in a sentence, $P(w|uv)$ is the trigram probability in the augmented LM, and $A_i$ represents the augmented corpus with $i^{th}$ sentence removed. This method of selection selects the sentences similar to the baseline corpus which results in maximum change in likelihood.

\subsubsection{Entropy Based Selection}
Itoh \etal~\cite{itoh2012n} employ the highest entropy-based selection of the N-best hypotheses for augmenting the acoustic model. We compute the sentence score based on the entropy of trigrams of Wikipedia with respect to the transcript language model as given in Equation \ref{equ:entropy}
\begin{equation}
\label{equ:entropy}
sentence\_score = \frac{\sum - P(w|T)\log P(w|T)}{\#(w)}
\end{equation}
where, $w$ is the trigram and $T$ is the baseline corpus.

\subsubsection{Random Selection}
The text selection methods for LM augmentation adapt various strategies with respect to the training text. In addition to these selection methods, we also conduct our experiments on a random selection of sentences from the Wikipedia text corpora for Telugu and Kannada and augment the corresponding language models. We randomly select 50\% of the Wikipedia text for augmenting the baseline language model.

\subsection{Different Sized Data Subsets}
\label{sect:datasize}
We study the effect of our minimal language model augmenting approach for initial decoding and later lattice rescoring on different-sized datasets. We compare the performance of our method with fully Wikipedia-augmented decoding on 10 hours, 20 hours, and 30 hours subsets of the Telugu speech dataset along with 4 hours of the Kannada dataset and 40 hours of the Telugu dataset.

\section{Results and Discussion}
\label{sect:result}
We discuss the results regarding different text selection methods in section \ref{sect:text_sel_res} followed by section \ref{sect:data_size_res} where we discuss the results concerning different sizes of datasets. The results presented in both sections comprise WER obtained and the effect of language model augmentation with various techniques on OOV and IV recognition. The results are listed as a percentage of OOV and IV words recognized.

\subsection{Text Selection based Language Model Augmentation}
\label{sect:text_sel_res}
This section depicts the results of different text selection methods for Telugu and Kannada ASR. We list the effect of language model augmentation on WER, OOV and IV recognition for both Telugu and Kannada with different text selection methods that have been employed. Also, the effect of applying our proposed method is depicted under the column named `Rescore after OWALM Decode'.

Table \ref{tab:telugu_textselection} and Table \ref{tab:kannada_textselection} list the WERs for different selection methods for Telugu and Kannada ASR respectively. Different subsets of Wikipedia text are used, based on the selection methods applied, to augment the baseline language model. The first column lists the selection method used. The second column lists the WERs obtained after decoding with the language models augmented using the corresponding selection methods. The third column lists the WER obtained when the lattices generated after initial decoding with the baseline language model, are rescored with the augmented language model. Finally, the fourth column depicts the effect of our method, which is to initially augment the baseline language model with OOT words, perform initial decode and then rescore the lattices with the corresponding augment language models. Likewise, Table \ref{tab:telugu_oov} and Table \ref{tab:kannada_oov} list the percentage of initial OOVs that were recognized while Table \ref{tab:telugu_iv} and Table \ref{tab:kannada_iv} list the percentage of initial in-vocabulary words that were recognized for different selection methods.

\begin{table*}[h!]
\caption{WER for Telugu based on different text selections based LM augmentation}
\label{tab:telugu_textselection}
\begin{tabular}{|l|ccc|}
\hline
&\multicolumn{3}{|c|}{\textbf {WER (\%)}}\\
\textbf{LM Augmentation} & \textbf{Decoding} & \textbf{Lattice Rescoring} & \textbf{Rescore after OWALM Decode}\\
&&&\textbf{(our method)}\\
\hline
Baseline & 26.76 & $-$ & $-$\\
Full Wiki (reference)& 21.01 & 25.78 & \textbf{20.92}\\
Contrastive Selection & 21.12 & 26.20 & \textbf{21.11}\\
Delta Likelihood Selection & 21.69 & 26.06 & \textbf{21.53}\\
Entropy-Based Selection & 21.69 & 26.11 & \textbf{21.56}\\
Random Selection & 21.47 & 25.90 & \textbf{21.27}\\
\hline
\end{tabular}
\tablenotes{$-$ Baseline WER is specified after including all the words from Wiki into the lexicon}
\tablenotes{$-$ OWALM : OOT Words Augmented Language Model}
\end{table*}

\begin{table*}[h!]
\caption{WER for Kannada based on different text selections based LM augmentation}
\label{tab:kannada_textselection}
\begin{tabular}{|l|ccc|}
\hline
&\multicolumn{3}{|c|}{\textbf {WER (\%)}}\\
\textbf{LM Augmentation} & \textbf{Decoding} & \textbf{Lattice Rescoring} & \textbf{Rescore after OWALM Decode}\\
&&&\textbf{(our method)}\\
\hline
Baseline & 52.01 & $-$ & $-$\\
Full Wiki (reference) & \textbf{28.12} & 50.08 & 30.27\\
Contrastive Selection & {\textbf{29.82}} & 50.95 & 31.67\\
Delta Likelihood Selection & {\textbf{28.42}} & 50.79 & 31.48\\
Entropy-Based Selection & {\textbf{29.77}} & 50.87 & 30.43\\
Random Selection & {\textbf{29.49}} & 50.89 & 31.09\\
\hline
\end{tabular}
\tablenotes{$-$ Baseline WER is specified after including all the words from Wiki into the lexicon}
\tablenotes{$-$ OWALM : OOT Words Augmented Language Model}
\end{table*}

We consider WER obtained by decoding using a language model augmented with complete Wikipedia text as the reference for the analysis. The best WER reduction observed for Telugu ASR, trained on a 40-hour corpus with a baseline LM of 44882 sentences, is 5.75\% absolute and 21.49\% relative using count merge base LM augmentation with full Wikipedia text. On the other hand, Kannada ASR which is trained only on 4 hours of speech with a baseline LM of 2647 sentences, achieves a best WER reduction of 23.89\% absolute and 45.93\% relative. This is more significant because of the very small baseline data.

It can be seen from Table \ref{tab:telugu_textselection} and Table \ref{tab:kannada_textselection} that selecting only 50 per cent of the Wikipedia text using either contrastive, delta likelihood, entropy-based for augmentation purposes and decoding results in a reduction of WER approximately as same as decoding with complete Wikipedia augmented language model. This is because the selection methods help select only meaningful sentences related to the baseline corpus. Wikipedia text contains garbage sentences as well which are eliminated through this selection. This helps in reducing the size of the language model for decoding. Interestingly, similar improvements in accuracy are obtained when the baseline language model is augmented with a random selection of sentences from Wikipedia text. This may be the case due to high OOV rates. A randomly selected subset containing sufficient OOV words is highly probable, thus improving performance. However, there is always uncertainty about how much to select, irrespective of the selection technique. Our approach eliminates this uncertainty by augmenting the baseline language model with the unigram counts of all the OOC words from the larger corpus for initial decoding. The lattices generated can then be rescored with a language model augmented with an entire large text corpus (refer to the entry for Full Wiki LM Augmentation in Table \ref{tab:telugu_textselection} and Table \ref{tab:kannada_textselection}) to obtain an effective reduction in WER.

Rescoring the lattices after initial decoding with the baseline language model leads to only marginal improvement in WER. This is because, in the case of low-resource languages, the lattices generated with baseline LM decoding may not contain many words due to the high OOV rate. As rescoring only updates the existing path probabilities based on new LM, only sub-words, if any, present in the lattice are recognized, and no new words are added. Hence, there is only a marginal improvement in WER. For example, the word `AADEESHAALAMERAKU' in Telugu ASR is correctly recognized after rescore because the sub-words `AADEESHAALA' and `MERAKU' are two sub-words present after baseline LM decode.

\begin{table*}[h!]
\caption{OOV Recovery in Telugu ASR based on different text selections based LM augmentation}
\label{tab:telugu_oov}
\begin{tabular}{|l|ccc|}
\hline
&\multicolumn{3}{|c|}{\textbf {OOV Recognized (\%)}}\\
\textbf{LM Augmentation} & \textbf{Decoding} & \textbf{Lattice Rescoring} & \textbf{Rescore after OWALM Decode}\\
&&&\textbf{(our method)}\\
\hline
Baseline & $-$ & $-$ & $-$\\
Full Wiki & \textbf{36.38} & 6.84 & 35.9\\
Contrastive Selection & 35.41 & 6.1 & \textbf{35.53}\\
Delta Likelihood Selection & 32.63 & 6.02 & \textbf{33}\\
Entropy-Based Selection & 31.72 & 5.72 & \textbf{31.87}\\
Random Selection & 33.41 & 6.69 & \textbf{32.20}\\
\hline
\end{tabular}
\tablenotes{$-$ OWALM : OOT Words Augmented Language Model}
\end{table*}

\begin{table*}[h!]
\caption{OOV Recovery in Kannada ASR based on different text selections based LM augmentation}
\label{tab:kannada_oov}
\begin{tabular}{|l|ccc|}
\hline
&\multicolumn{3}{|c|}{\textbf {OOV Recognized (\%)}}\\
\textbf{LM Augmentation} & \textbf{Decoding} & \textbf{Lattice Rescoring} & \textbf{Rescore after OWALM Decode}\\
&&&\textbf{(our method)}\\
\hline
Baseline & $-$ & $-$ & $-$\\
Full Wiki & \textbf{62.28} & 6.72 & 56.54\\
Contrastive Selection & {\textbf{56.31}} & 6.72 & 53.96\\
Delta Likelihood Selection & {\textbf{57.12}} & 6.03 & 54.76\\
Entropy-Based Selection & {\textbf{59.53}} & 6.37 & 55.74\\
Random Selection & {\textbf{58.55}} & 6.43 & 54.13\\
\hline
\end{tabular}
\tablenotes{$-$ OWALM : OOT Words Augmented Language Model}
\end{table*}

According to our method, augmenting the baseline language model with a minimum of only OOV words with respect to the baseline for decoding generates lattices that contain these words and hence rescoring such a lattice with a larger language model results in a significant reduction of WER as is evident from Table \ref{tab:telugu_textselection} and Table \ref{tab:kannada_textselection}. The WER obtained for different selection methods using our method is consistently very close to the reference WER. It is better for Telugu and slightly more for Kannada. Also, from Table \ref{tab:telugu_oov} and Table \ref{tab:kannada_oov}, it is seen that the percentage of out-of-vocabulary words recognized using our method improves and is very close to the reference. From Table \ref{tab:telugu_iv} and Table \ref{tab:kannada_iv}, we see that the percentage of in-vocabulary words recognized also improves with respect to the baseline.

\begin{table*}[h!]
\caption{IV Recognition in Telugu ASR based on different text selections based LM augmentation}
\label{tab:telugu_iv}
\begin{tabular}{|l|ccc|}
\hline
&\multicolumn{3}{|c|}{\textbf {IV Recognized (\%)}}\\
\textbf{LM Augmentation} & \textbf{Decoding} & \textbf{Lattice Rescoring} & \textbf{Rescore after OWALM Decode}\\
&&&\textbf{(our method)}\\
\hline
Baseline & $91.97$ & $-$ & $-$\\
Full Wiki & \textbf{94.59} & 6.84 & 35.9\\
Contrastive Selection & \textbf{94.34} & 92.87 & \textbf{94.34}\\
Delta Likelihood Selection & \textbf{94.26} & 92.57 & 94.06\\
Entropy-Based Selection & \textbf{94.28} & 92.28 & \textbf{94.28}\\
Random Selection & \textbf{94.17} & 92.87 & 92.85\\
\hline
\end{tabular}
\tablenotes{$-$ OWALM : OOT Words Augmented Language Model}
\end{table*}

\begin{table*}[h!]
\caption{IV Recognition in Kannada ASR based on different text selections based LM augmentation}
\label{tab:kannada_iv}
\begin{tabular}{|l|ccc|}
\hline
&\multicolumn{3}{|c|}{\textbf {IV Recognized (\%)}}\\
\textbf{LM Augmentation} & \textbf{Decoding} & \textbf{Lattice Rescoring} & \textbf{Rescore after OWALM Decode}\\
&&&\textbf{(our method)}\\
\hline
Baseline & 64.83 & $-$ & $-$\\
Full Wiki & \textbf{78.99} & 66.12 & 76.98\\
Contrastive Selection & {\textbf{77.21}} & 67.17 & 77.19\\
Delta Likelihood Selection & 73.01 & 67.72 & \textbf{76.17}\\
Entropy-Based Selection & {\textbf{78.84}} & 66.78 & 78.4\\
Random Selection & {\textbf{77.59}} & 67.65 & 75.08\\
\hline
\end{tabular}
\tablenotes{$-$ OWALM : OOT Words Augmented Language Model}
\end{table*}

\begin{figure}[!t]
\centering{
\includegraphics[width=0.95\columnwidth]{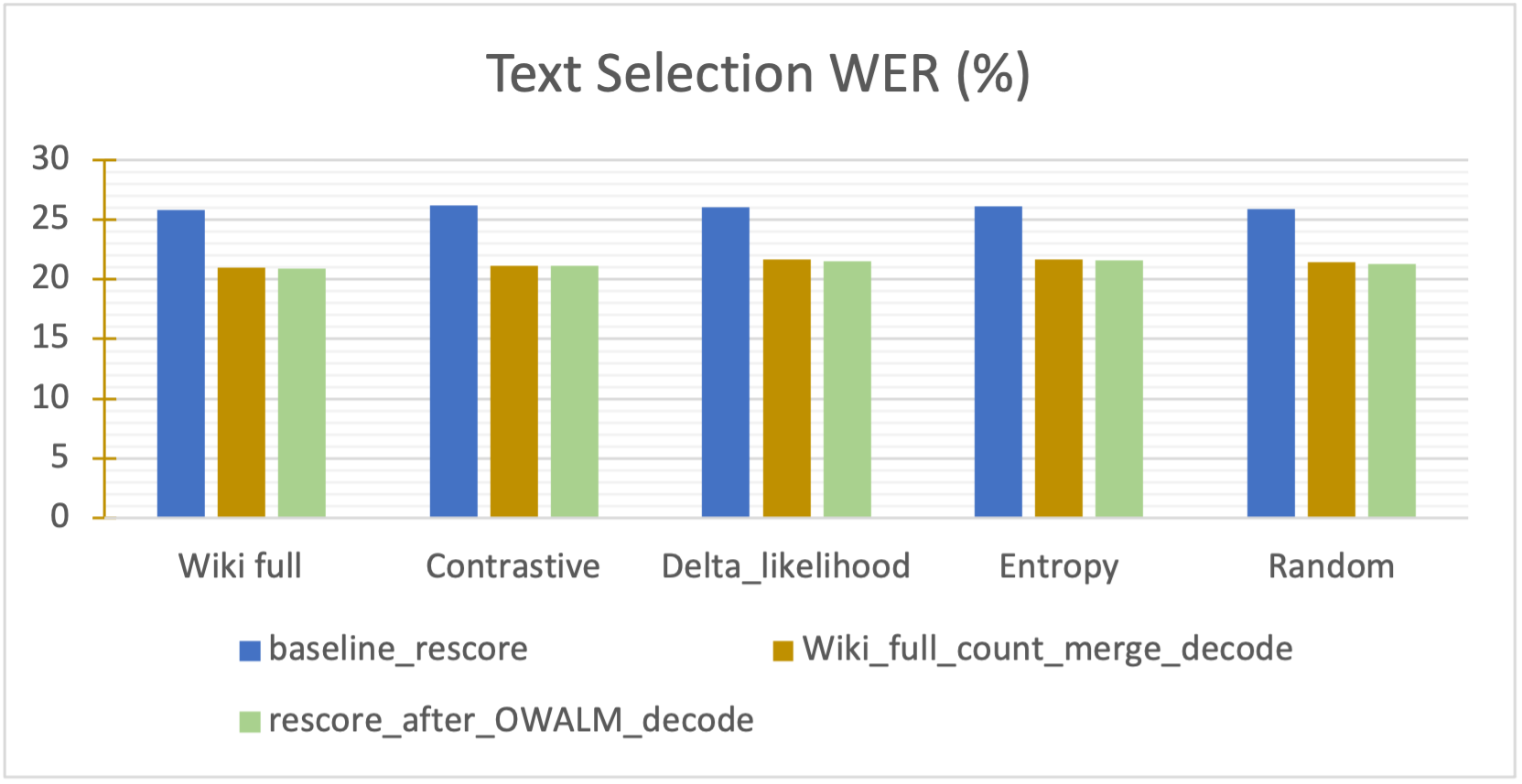}
}
\caption{WER for different selection methods in Telugu}
\label{fig:tel_text_selection}
\end{figure}

\begin{figure}[!t]
\centering{
\includegraphics[width=0.95\columnwidth]{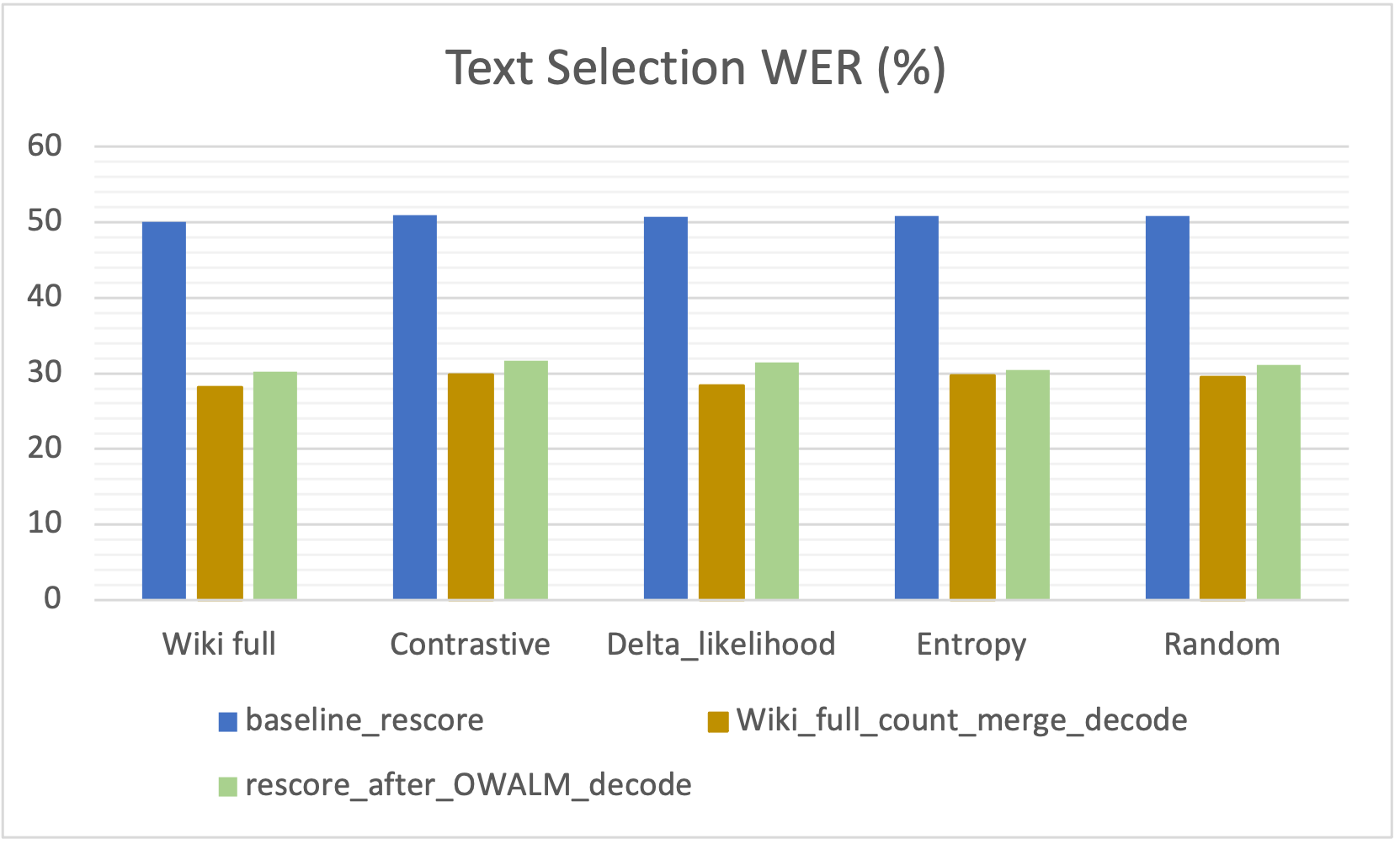}
}
\caption{WER for different selection methods in Kannada}
\label{fig:kan_text_selection}
\end{figure}

\begin{sloppypar}
Figure \ref{fig:tel_text_selection} and Figure \ref{fig:kan_text_selection} depict the comparison between augmented language model rescore with lattices generated from baseline decode (blue bar), decode with augmented language model (brown bar) and our proposed method (green). Y-axis in the figure represents WER as a percentage. The WER obtained after the initial baseline LM decoding followed by rescoring with a bigger language model shows only a marginal reduction from the baseline WER for both Kannada (52.01\%) and Telugu (26.76\%) datasets. The results show that the rescoring method proposed in this paper, `rescore\_after\_OWALM\_decode' (green bar), results in a WER that is comparable with decoding using a bigger language model (brown bar), and this is true for all selection methods. Our proposed method is more effective in leveraging the availability of a bigger text corpus for low-resource languages and also saves computational resources.
\end{sloppypar}

\subsection{Datasets of Different Sizes}
\label{sect:data_size_res}
In this section, the results depict the effect of initial decoding with baseline LM enhanced to include OOT unigrams and then performing lattice rescoring with LM augmented with full Wikipedia text on different-sized datasets as listed in Table \ref{tab:data_size}. 

\begin{table*}[h!]
\caption{Results for different sized datasets}
\label{tab:data_size}
\begin{tabular}{|l|l|ccc|}
\hline
\textbf{Duration} & \textbf{Language Model} & \textbf{WER (\%)} & \textbf{OOV Recognized (\%)} & \textbf{IV Recognized (\%)}\\
 & (Decode / Rescore) & & &\\
\hline
\multirow{4}{*} {4hrs (Kannada)} & Baseline & 52.01 & $-$ & 68.43\\
& OWALM Decode & 38.79 & 35.94 & 69.30\\
& Wiki Decode & \textbf{28.12} & \textbf{62.28} & \textbf{78.99}\\
& Wiki Rescore after OWALM Decode & 30.27 & 56.54 & 76.98\\
\hline
\hline
\multirow{4}{*} {10hrs (Telugu)} & Baseline & 43.43 & $-$ & 86.89\\
& OWALM Decode & 33.81 & 28.75 & 85.2\\
& Wiki Decode & \textbf{29.16} & \textbf{34.38} & \textbf{91.47}\\
& Wiki Rescore after OWALM Decode & 29.44 & 34.32 & 90.77\\
\hline
\hline
\multirow{4}{*} {20hrs (Telugu)} & Baseline & 29.97 & $-$ & 91.53\\
& OWALM Decode & 24.52 & 31.33 & 91.06\\
& Wiki Decode & 22.53 & 34.9 & 93.74\\
& Wiki Rescore after OWALM Decode & \textbf{22.3} & \textbf{36.5} & \textbf{94.06}\\
\hline
\hline
\multirow{4}{*} {30hrs (Telugu)} & Baseline & 27.46 & $-$ & 91.64\\
& OWALM Decode & 22.86 & 32.14 & 92.18\\
& Wiki Decode & 20.95 & \textbf{36.83} & \textbf{94.69}\\
& Wiki Rescore after OWALM Decode & \textbf{20.89} & 36.5 & 94.27\\
\hline
\hline
\multirow{4}{*} {40hrs (Telugu)} & Baseline & 26.76 & $-$ & 91.97\\
& OWALM Decode & 22.66 & 31.27 & 92.05\\
& Wiki Decode & 21.01 & \textbf{36.38} & \textbf{94.59}\\
& Wiki Rescore after OWALM Decode & \textbf{20.92} & 35.90 & 94.27\\
\hline
\end{tabular}
\tablenotes{$-$ OWALM - OOT Words Augmented Language Model}
\end{table*}

The relative improvement obtained after language model augmentation is more pronounced for smaller speech datasets and reduces with an increase in the size of the corpora. Nevertheless, the results obtained from our approach are yet comparable with decoding using a full Wikipedia augmented language model across all sized datasets. Our method shows more improvement in the full Wikipedia decoding as the size of the dataset improves.

\begin{figure}[!t]
\centering{
\includegraphics[width=0.95\columnwidth]{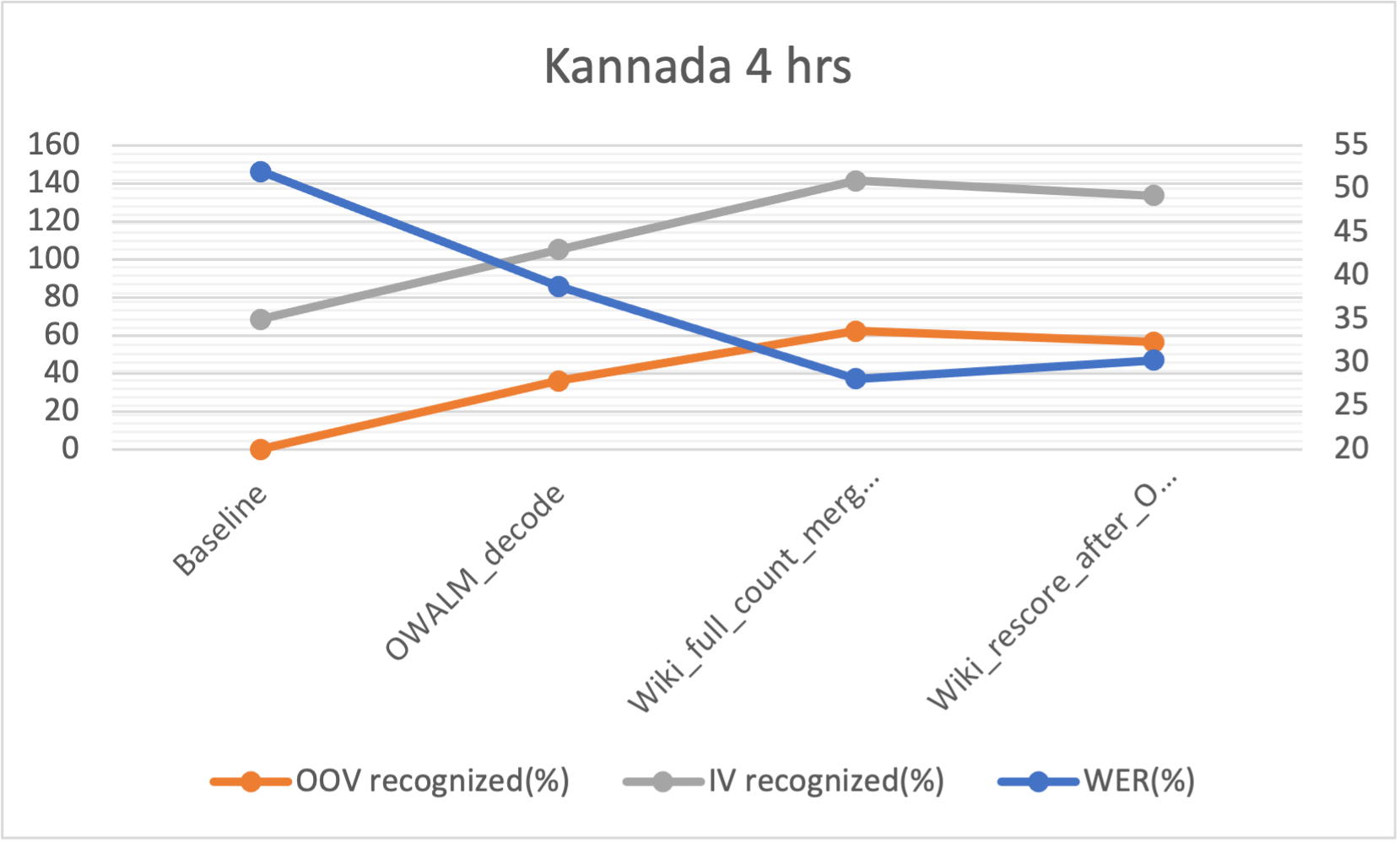}
}
\caption{WER, OOV and IV percentages for 4 hours dataset}
\label{fig:4hr}
\end{figure}

\begin{figure}[!t]
\centering{
\includegraphics[width=0.95\columnwidth]{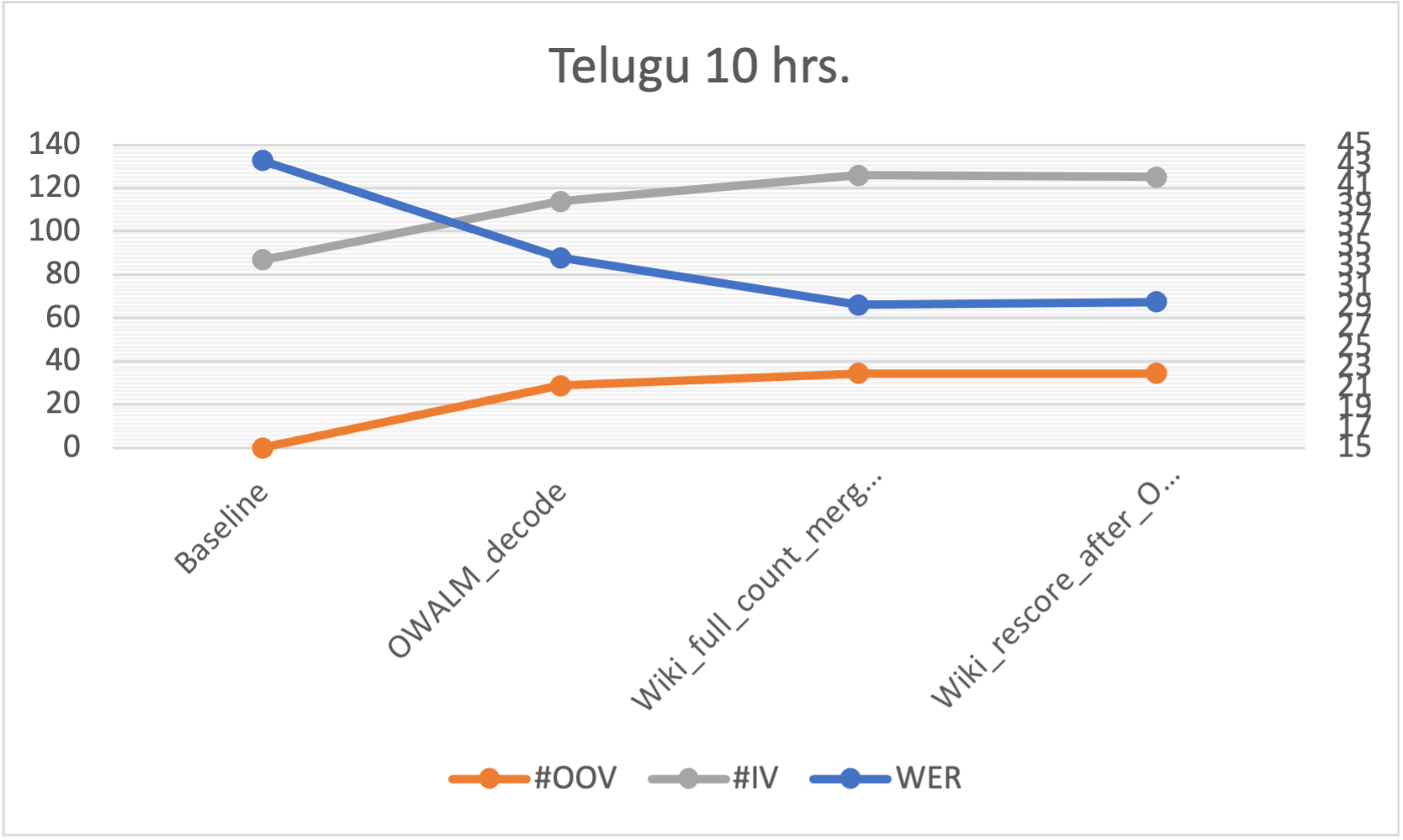}
}
\caption{WER, OOV and IV percentages for 10 hours dataset}
\label{fig:10hr}
\end{figure}

\begin{figure}[!t]
\centering{
\includegraphics[width=0.95\columnwidth]{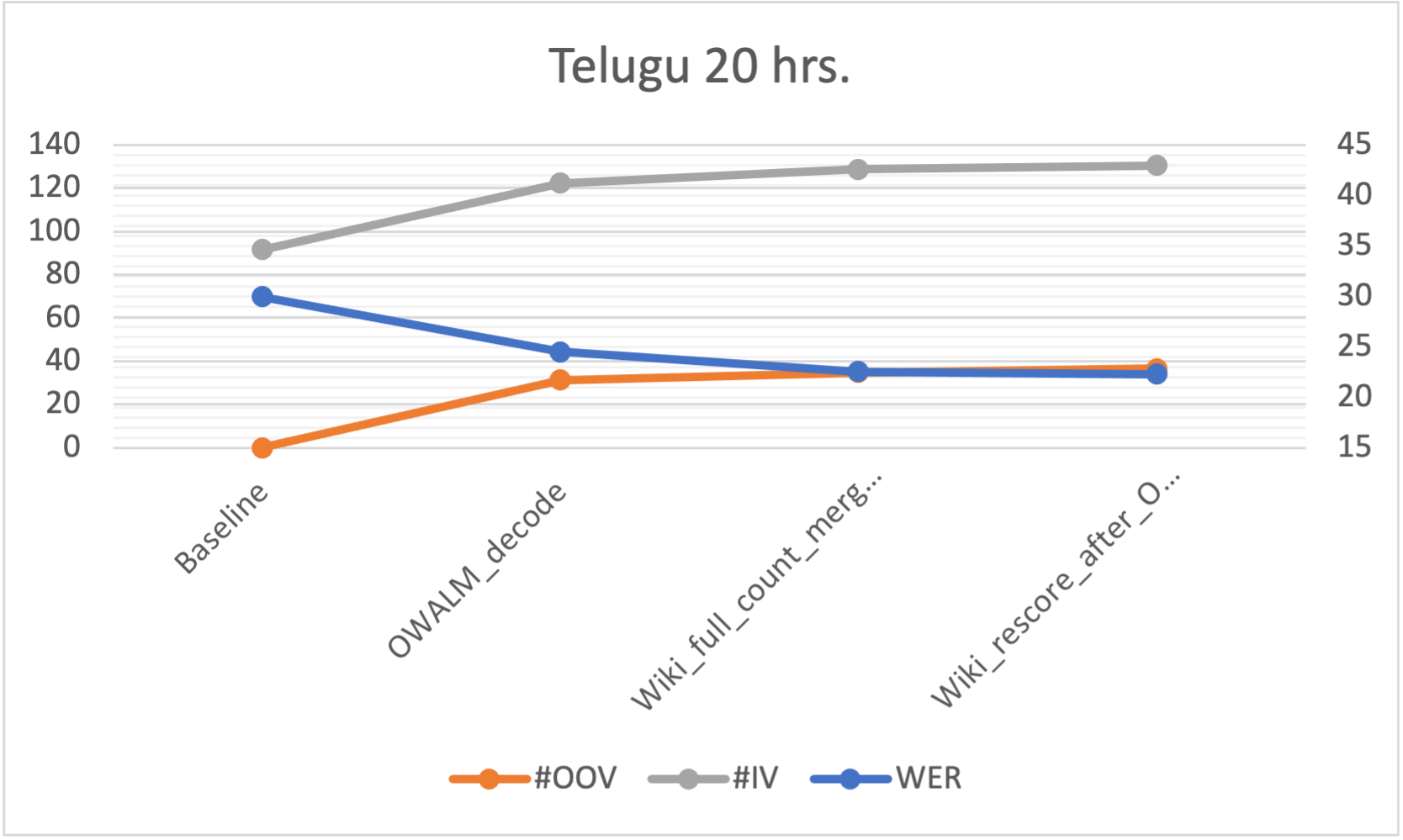}
}
\caption{WER, OOV and IV percentages for 20 hours dataset}
\label{fig:20hr}
\end{figure}

\begin{figure}[!t]
\centering{
\includegraphics[width=0.95\columnwidth]{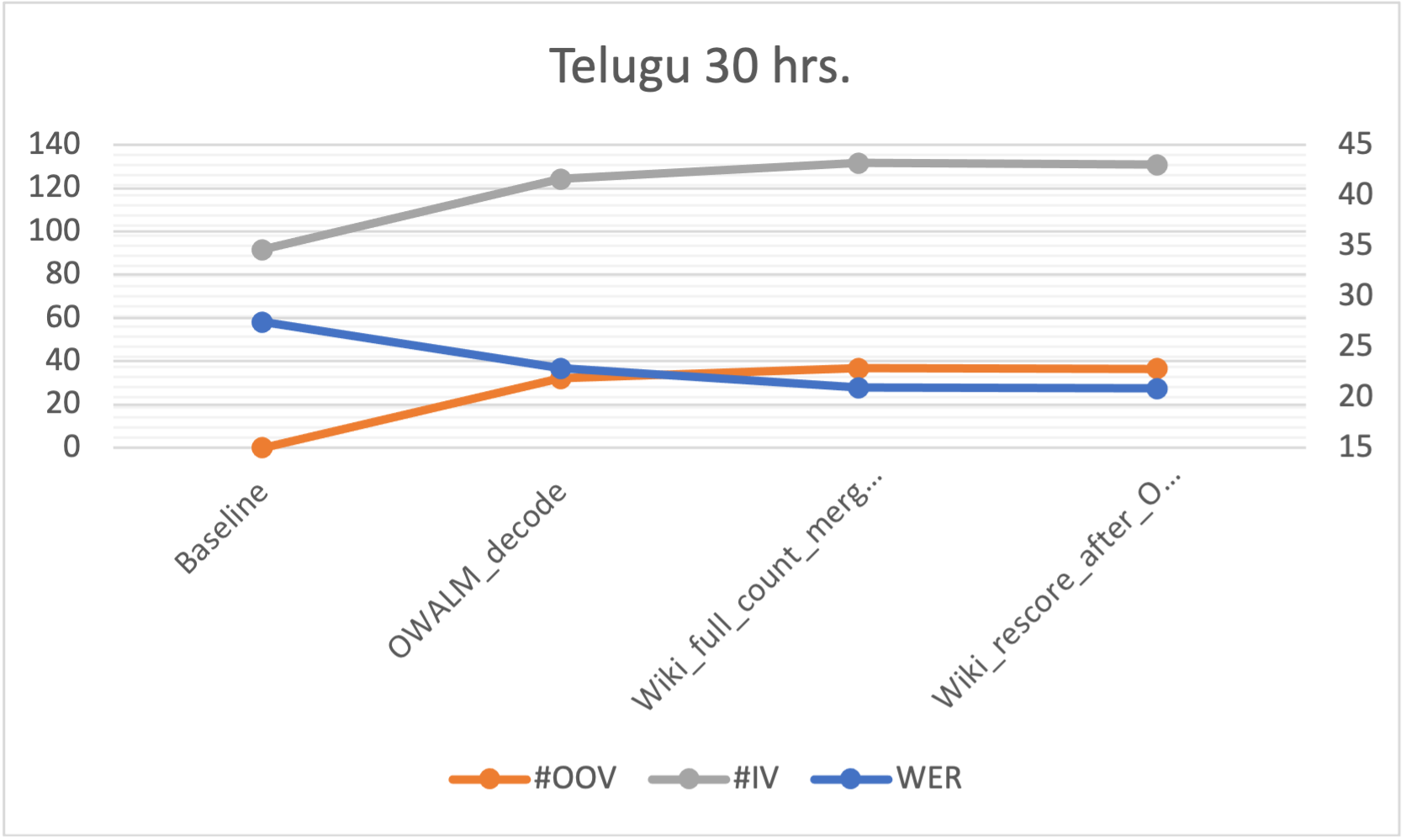}
}
\caption{WER, OOV and IV percentages for 30 hours dataset}
\label{fig:30hr}
\end{figure}

\begin{figure}[!t]
\centering{
\includegraphics[width=0.95\columnwidth]{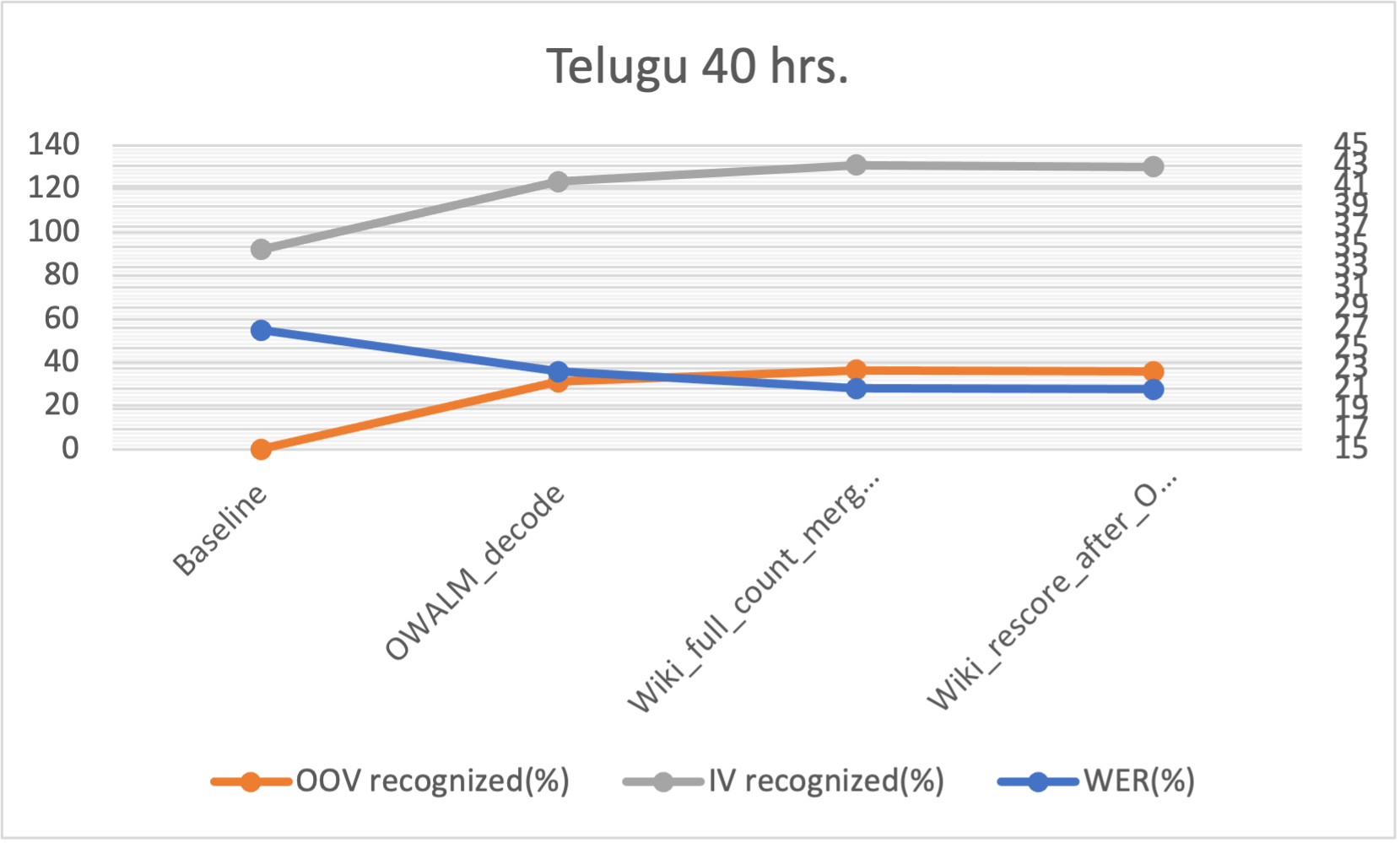}
}
\caption{WER, OOV and IV percentages for 40 hours dataset}
\label{fig:40hr}
\end{figure}

Figure \ref{fig:4hr}, Figure \ref{fig:10hr}, Figure \ref{fig:20hr}, Figure \ref{fig:30hr}, and Figure \ref{fig:40hr} depict the effect on WER, OOV and IV recognition for different sizes of the datasets – 4 hours of Kannada, 10 hours,  20 hours,  30 hours and 40 hours of Telugu respectively.  

\begin{sloppypar}
The data points for each of the figures are baseline (with no LM augmentation), OWALM decode, `Wiki\_full\_count\_merge\_decode' (decode with LM augmentation with complete Wikipedia using count merge) and `Wiki\_rescore\_after\_OWALM\_decode' (rescore with count merge augmented LM after decoding with OWALM). The data points for `oov\_wiki\_aug' and `oov\_words\_wiki\_aug\_rescore' are closer to each other when compared to the relative difference with the baseline.
\end{sloppypar}
This indicates that first decoding with baseline LM augmented with only OOT unigrams increases the language model scores for these OOV words. This makes later rescoring with a larger LM effective, and the accuracy is comparable with that obtained after decoding with a larger language model. Also, the IV recognition shows improvement with respect to the baseline. This is true across datasets of different sizes. The relative improvement, in terms of WER and OOV recovery, is, however, more pronounced for smaller datasets.

\section{Conclusions and Future Work}
\label{sect:conclusion}
Our experiments show that rescoring with a language model trained on larger text may not be very effective in the case of low-resource languages since the baseline language model is not sufficiently large to cover all the possible contexts for words in the vocabulary. Hence, it is essential that we augment the baseline language model with minimal data that comprises the enhanced vocabulary, obtain a lattice from the decoding graph built using the minimally enhanced grammar and then rescore with a larger language model. Our method of building a decoding graph from a grammar augmented with only out-of-train word unigrams from Wikipedia and then rescoring the obtained lattice with a larger language model is as effective (more effective in the case of Telugu and comparable in the case of Kannada) as decoding with an HCLG graph built using a language model augmented with the entire Wikipedia text. This method is applicable across different-sized datasets and also for different text selection methods. We can, thus, leverage the availability of larger amounts of non-domain text corpora while at the same time, reducing the computational overhead of decoding with a bigger language model. It would be interesting to investigate the application of our approach to other low-resource languages. Further, we also intend to explore our approach with the complementary morpheme-based approach and with approaches for named entity recognition. We leave this for future work.

\section*{Acknowledgements}
We thank Dr K.V. Subramanian, Head, Center for Cloud Computing and Big Data, PES University, Bangalore, for all the support.
\end{sloppypar}
\balance

\bibliography{mybibfile}

\begin{thebibliography}{10}

\bibitem{hazen2009query}
Hazen Timothy~J, Shen Wade, and White Christopher.
\newblock Query-by-example spoken term detection using phonetic posteriorgram
  templates.
\newblock In {\em 2009 IEEE Workshop on Automatic Speech Recognition \&
  Understanding}, pages 421--426. IEEE, 2009.

\bibitem{novotney2010cheap}
Novotney Scott and Callison-Burch Chris.
\newblock Cheap, fast and good enough: Automatic speech recognition with
  non-expert transcription.
\newblock In {\em Human Language Technologies: The 2010 Annual Conference of
  the North American Chapter of the Association for Computational Linguistics},
  pages 207--215, 2010.

\bibitem{thomas2010cross}
Thomas Samuel, Ganapathy Sriram, and Hermansky Hynek.
\newblock Cross-lingual and multi-stream posterior features for low resource
  lvcsr systems.
\newblock In {\em Eleventh Annual Conference of the International Speech
  Communication Association}, 2010.

\bibitem{bulyko2007web}
Bulyko Ivan, Ostendorf Mari, Siu Manhung, Ng~Tim, Stolcke Andreas, and
  {\c{C}}etin {\"O}zg{\"u}r.
\newblock Web resources for language modeling in conversational speech
  recognition.
\newblock {\em ACM Transactions on Speech and Language Processing (TSLP)},
  5(1):1--25, 2007.

\bibitem{mendels2015improving}
Mendels Gideon, Cooper Erica, Soto Victor, Hirschberg Julia, Gales Mark~JF,
  Knill Kate~M, Ragni Anton, and Wang Haipeng.
\newblock Improving speech recognition and keyword search for low resource
  languages using web data.
\newblock In {\em INTERSPEECH 2015: 16th Annual Conference of the International
  Speech Communication Association}, pages 829--833. International Speech
  Communication Association (ISCA), 2015.

\bibitem{sethy2006text}
Sethy Abhinav, Georgiou Panayiotis, and Narayanan Shrikanth.
\newblock Text data acquisition for domain-specific language models.
\newblock In {\em Proceedings of the 2006 Conference on Empirical Methods in
  Natural Language Processing}, pages 382--389, 2006.

\bibitem{parada2010spoken}
Parada Carolina, Sethy Abhinav, Dredze Mark, and Jelinek Frederick.
\newblock A spoken term detection framework for recovering out-of-vocabulary
  words using the web.
\newblock In {\em Eleventh annual conference of the international speech
  communication association}, 2010.

\bibitem{ng2005web}
Ng~Tim, Ostendorf Mari, Hwang Mei-Yuh, Siu Manhung, Bulyko Ivan, and Lei Xin.
\newblock Web-data augmented language models for mandarin conversational speech
  recognition.
\newblock In {\em Proceedings.(ICASSP'05). IEEE International Conference on
  Acoustics, Speech, and Signal Processing, 2005.}, volume~1, pages I--589.
  IEEE, 2005.

\bibitem{beck2019lstm}
Beck Eugen, Zhou Wei, Schl{\"u}ter Ralf, and Ney Hermann.
\newblock Lstm language models for lvcsr in first-pass decoding and
  lattice-rescoring.
\newblock {\em arXiv preprint arXiv:1907.01030}, 2019.

\bibitem{song2020specswap}
Song Xingcheng, Wu~Zhiyong, Huang Yiheng, Su~Dan, and Meng Helen.
\newblock Specswap: A simple data augmentation method for end-to-end speech
  recognition.
\newblock In {\em INTERSPEECH}, pages 581--585, 2020.

\bibitem{ko2015audio}
Ko~Tom, Peddinti Vijayaditya, Povey Daniel, and Khudanpur Sanjeev.
\newblock Audio augmentation for speech recognition.
\newblock In {\em Sixteenth annual conference of the international speech
  communication association}, 2015.

\bibitem{vachhani2018data}
Vachhani Bhavik, Bhat Chitralekha, and Kopparapu Sunil~Kumar.
\newblock Data augmentation using healthy speech for dysarthric speech
  recognition.
\newblock In {\em Interspeech}, pages 471--475, 2018.

\bibitem{tuske2014data}
T{\"u}ske Zolt{\'a}n, Golik Pavel, Nolden David, Schl{\"u}ter Ralf, and Ney
  Hermann.
\newblock Data augmentation, feature combination, and multilingual neural
  networks to improve asr and kws performance for low-resource languages.
\newblock In {\em Fifteenth Annual Conference of the International Speech
  Communication Association}, 2014.

\bibitem{jaitly2013vocal}
Jaitly Navdeep and Hinton Geoffrey~E.
\newblock Vocal tract length perturbation (vtlp) improves speech recognition.
\newblock In {\em Proc. ICML Workshop on Deep Learning for Audio, Speech and
  Language}, volume 117, page~21, 2013.

\bibitem{chen2020data}
Chen Guoguo, Na~Xingyu, Wang Yongqing, Yan Zhiyong, Zhang Junbo, Ma~Sifan, and
  Wang Yujun.
\newblock Data augmentation for children's speech recognition--the" ethiopian"
  system for the slt 2021 children speech recognition challenge.
\newblock {\em arXiv preprint arXiv:2011.04547}, 2020.

\bibitem{medennikov2018investigation}
Medennikov Ivan, Khokhlov Yuri~Y, Romanenko Aleksei, Popov Dmitry, Tomashenko
  Natalia~A, Sorokin Ivan, and Zatvornitskiy Alexander.
\newblock An investigation of mixup training strategies for acoustic models in
  asr.
\newblock In {\em Interspeech}, pages 2903--2907, 2018.

\bibitem{saon2019sequence}
Saon George, T{\"u}ske Zolt{\'a}n, Audhkhasi Kartik, and Kingsbury Brian.
\newblock Sequence noise injected training for end-to-end speech recognition.
\newblock In {\em ICASSP 2019-2019 IEEE International Conference on Acoustics,
  Speech and Signal Processing (ICASSP)}, pages 6261--6265. IEEE, 2019.

\bibitem{zhu2019mixup}
Zhu Yingke, Ko~Tom, and Mak Brian.
\newblock Mixup learning strategies for text-independent speaker verification.
\newblock In {\em Interspeech}, pages 4345--4349, 2019.

\bibitem{ghahremani2017investigation}
Ghahremani Pegah, Manohar Vimal, Hadian Hossein, Povey Daniel, and Khudanpur
  Sanjeev.
\newblock Investigation of transfer learning for asr using lf-mmi trained
  neural networks.
\newblock In {\em 2017 IEEE Automatic Speech Recognition and Understanding
  Workshop (ASRU)}, pages 279--286. IEEE, 2017.

\bibitem{manohar2017jhu}
Manohar Vimal, Povey Daniel, and Khudanpur Sanjeev.
\newblock Jhu kaldi system for arabic mgb-3 asr challenge using diarization,
  audio-transcript alignment and transfer learning.
\newblock In {\em 2017 IEEE Automatic Speech Recognition and Understanding
  Workshop (ASRU)}, pages 346--352. IEEE, 2017.

\bibitem{wang2020cross}
Wang Sicheng, Li~Wei, Siniscalchi Sabato~Marco, and Lee Chin-Hui.
\newblock A cross-task transfer learning approach to adapting deep speech
  enhancement models to unseen background noise using paired senone
  classifiers.
\newblock In {\em ICASSP 2020-2020 IEEE International Conference on Acoustics,
  Speech and Signal Processing (ICASSP)}, pages 6219--6223. IEEE, 2020.

\bibitem{li2018training}
Li~Jason, Gadde Ravi, Ginsburg Boris, and Lavrukhin Vitaly.
\newblock Training neural speech recognition systems with synthetic speech
  augmentation.
\newblock {\em arXiv preprint arXiv:1811.00707}, 2018.

\bibitem{laptev2020you}
Laptev Aleksandr, Korostik Roman, Svischev Aleksey, Andrusenko Andrei,
  Medennikov Ivan, and Rybin Sergey.
\newblock You do not need more data: Improving end-to-end speech recognition by
  text-to-speech data augmentation.
\newblock In {\em 2020 13th International Congress on Image and Signal
  Processing, BioMedical Engineering and Informatics (CISP-BMEI)}, pages
  439--444. IEEE, 2020.

\bibitem{rosenberg2019speech}
Rosenberg Andrew, Zhang Yu, Ramabhadran Bhuvana, Jia Ye, Moreno Pedro,
  Wu~Yonghui, and Wu~Zelin.
\newblock Speech recognition with augmented synthesized speech.
\newblock In {\em 2019 IEEE automatic speech recognition and understanding
  workshop (ASRU)}, pages 996--1002. IEEE, 2019.

\bibitem{benevs2020text}
Bene{\v{s}} Karel and Burget Luk{\'a}{\v{s}}.
\newblock Text augmentation for language models in high error recognition
  scenario.
\newblock {\em arXiv preprint arXiv:2011.06056}, 2020.

\bibitem{peyser2020improving}
Peyser Cal, Mavandadi Sepand, Sainath Tara~N, Apfel James, Pang Ruoming, and
  Kumar Shankar.
\newblock Improving tail performance of a deliberation e2e asr model using a
  large text corpus.
\newblock {\em arXiv preprint arXiv:2008.10491}, 2020.

\bibitem{sharma2020improving}
Sharma Yash, Abraham Basil, Taneja Karan, and Jyothi Preethi.
\newblock Improving low resource code-switched asr using augmented
  code-switched tts.
\newblock {\em arXiv preprint arXiv:2010.05549}, 2020.

\bibitem{meng2021mixspeech}
Meng Linghui, Xu~Jin, Tan Xu, Wang Jindong, Qin Tao, and Xu~Bo.
\newblock Mixspeech: Data augmentation for low-resource automatic speech
  recognition.
\newblock In {\em ICASSP 2021-2021 IEEE International Conference on Acoustics,
  Speech and Signal Processing (ICASSP)}, pages 7008--7012. IEEE, 2021.

\bibitem{rossenbach2020generating}
Rossenbach Nick, Zeyer Albert, Schl{\"u}ter Ralf, and Ney Hermann.
\newblock Generating synthetic audio data for attention-based speech
  recognition systems.
\newblock In {\em ICASSP 2020-2020 IEEE International Conference on Acoustics,
  Speech and Signal Processing (ICASSP)}, pages 7069--7073. IEEE, 2020.

\bibitem{lin2020training}
Lin James, Kilgour Kevin, Roblek Dominik, and Sharifi Matthew.
\newblock Training keyword spotters with limited and synthesized speech data.
\newblock In {\em ICASSP 2020-2020 IEEE International Conference on Acoustics,
  Speech and Signal Processing (ICASSP)}, pages 7474--7478. IEEE, 2020.

\bibitem{hartmann2016two}
Hartmann William, Ng~Tim, Hsiao Roger, Tsakalidis Stavros, and Schwartz
  Richard~M.
\newblock Two-stage data augmentation for low-resourced speech recognition.
\newblock In {\em Interspeech}, pages 2378--2382, 2016.

\bibitem{hailu2020improving}
Hailu Nirayo, Siegert Ingo, and N{\"u}rnberger Andreas.
\newblock Improving automatic speech recognition utilizing audio-codecs for
  data augmentation.
\newblock In {\em 2020 IEEE 22nd International Workshop on Multimedia Signal
  Processing (MMSP)}, pages 1--5. IEEE, 2020.

\bibitem{rebai2017improving}
Rebai Ilyes, BenAyed Yessine, Mahdi Walid, and Lorr{\'e} Jean-Pierre.
\newblock Improving speech recognition using data augmentation and acoustic
  model fusion.
\newblock {\em Procedia Computer Science}, 112:316--322, 2017.

\bibitem{rosenberg2017end}
Rosenberg Andrew, Audhkhasi Kartik, Sethy Abhinav, Ramabhadran Bhuvana, and
  Picheny Michael.
\newblock End-to-end speech recognition and keyword search on low-resource
  languages.
\newblock In {\em 2017 IEEE International Conference on Acoustics, Speech and
  Signal Processing (ICASSP)}, pages 5280--5284. IEEE, 2017.

\bibitem{vydana2018exploration}
Vydana Hari~Krishna, Gurugubelli Krishna, Vegesna Vishnu Vidyadhara~Raju, and
  Vuppala Anil~Kumar.
\newblock An exploration towards joint acoustic modeling for indian languages:
  Iiit-h submission for low resource speech recognition challenge for indian
  languages, interspeech 2018.
\newblock In {\em INTERSPEECH}, pages 3192--3196, 2018.

\bibitem{fathima2018tdnn}
Fathima Noor, Patel Tanvina, Mahima C, and Iyengar Anuroop.
\newblock Tdnn-based multilingual speech recognition system for low resource
  indian languages.
\newblock In {\em INTERSPEECH}, pages 3197--3201, 2018.

\bibitem{pulugundla2018but}
Pulugundla Bhargav, Baskar Murali~Karthick, Kesiraju Santosh, Egorova
  Ekaterina, Karafi{\'a}t Martin, Burget Luk{\'a}s, and Cernock{\`y} Jan.
\newblock But system for low resource indian language asr.
\newblock In {\em INTERSPEECH}, pages 3182--3186, 2018.

\bibitem{shetty2018articulatory}
Shetty Vishwas~M, Sharon Rini~A, Abraham Basil, Seeram Tejaswi, Prakash Anusha,
  Ravi Nithya, and Umesh Srinivasan.
\newblock Articulatory and stacked bottleneck features for low resource speech
  recognition.
\newblock In {\em INTERSPEECH}, pages 3202--3206, 2018.

\bibitem{billa2018isi}
Billa Jayadev.
\newblock Isi asr system for the low resource speech recognition challenge for
  indian languages.
\newblock In {\em INTERSPEECH}, pages 3207--3211, 2018.

\bibitem{yilmaz2018acoustic}
Y{\i}lmaz Emre, Heuvel Henk van~den, and Leeuwen David~Avan.
\newblock Acoustic and textual data augmentation for improved asr of
  code-switching speech.
\newblock {\em arXiv preprint arXiv:1807.10945}, 2018.

\bibitem{klakow1999oov}
Klakow Dietrich, Rose Georg, and Aubert Xavier.
\newblock Oov-detection in large vocabulary system using automatically defined
  word-fragments as fillers.
\newblock In {\em Sixth European Conference on Speech Communication and
  Technology}, 1999.

\bibitem{schaaf2001detection}
Schaaf Thomas.
\newblock Detection of oov words using generalized word models and a semantic
  class language model.
\newblock In {\em Seventh European Conference on Speech Communication and
  Technology}, 2001.

\bibitem{kitaoka2021dynamic}
Kitaoka Norihide, Chen Bohan, and Obashi Yuya.
\newblock Dynamic out-of-vocabulary word registration to language model for
  speech recognition.
\newblock {\em EURASIP Journal on Audio, Speech, and Music Processing},
  2021(1):1--8, 2021.

\bibitem{thomas2019detection}
Thomas Samuel, Audhkhasi Kartik, T{\"u}ske Zolt{\'a}n, Huang Yinghui, and
  Picheny Michael.
\newblock Detection and recovery of oovs for improved english broadcast news
  captioning.
\newblock In {\em INTERSPEECH}, pages 2973--2977, 2019.

\bibitem{hazen2001comparison}
Hazen Timothy~J and Bazzi Issam.
\newblock A comparison and combination of methods for oov word detection and
  word confidence scoring.
\newblock In {\em 2001 IEEE International Conference on Acoustics, Speech, and
  Signal Processing. Proceedings (Cat. No. 01CH37221)}, volume~1, pages
  397--400. IEEE, 2001.

\bibitem{yazgan2004hybrid}
Yazgan Ali and Saraclar Murat.
\newblock Hybrid language models for out of vocabulary word detection in large
  vocabulary conversational speech recognition.
\newblock In {\em 2004 IEEE International Conference on Acoustics, Speech, and
  Signal Processing}, volume~1, pages I--745. IEEE, 2004.

\bibitem{ketabdar2007detection}
Ketabdar Hamed, Hannemann Mirko, and Hermansky Hynek.
\newblock Detection of out-of-vocabulary words in posterior based asr.
\newblock In {\em Eighth Annual Conference of the International Speech
  Communication Association}, 2007.

\bibitem{white2008confidence}
White Christopher, Zweig Geoffrey, Burget Lukas, Schwarz Petr, and Hermansky
  Hynek.
\newblock Confidence estimation, oov detection and language id using
  phone-to-word transduction and phone-level alignments.
\newblock In {\em 2008 IEEE International Conference on Acoustics, Speech and
  Signal Processing}, pages 4085--4088. IEEE, 2008.

\bibitem{rastrow2009new}
Rastrow Ariya, Sethy Abhinav, and Ramabhadran Bhuvana.
\newblock A new method for oov detection using hybrid word/fragment system.
\newblock In {\em 2009 IEEE International Conference on Acoustics, Speech and
  Signal Processing}, pages 3953--3956. IEEE, 2009.

\bibitem{zhang2020oov}
Zhang Xiaohui, Povey Daniel, and Khudanpur Sanjeev.
\newblock Oov recovery with efficient 2nd pass decoding and open-vocabulary
  word-level rnnlm rescoring for hybrid asr.
\newblock In {\em ICASSP 2020-2020 IEEE International Conference on Acoustics,
  Speech and Signal Processing (ICASSP)}, pages 6334--6338. IEEE, 2020.

\bibitem{lin2007oov}
Lin Hui, Bilmes Jeff, Vergyri Dimitra, and Kirchhoff Katrin.
\newblock Oov detection by joint word/phone lattice alignment.
\newblock In {\em 2007 IEEE Workshop on Automatic Speech Recognition \&
  Understanding (ASRU)}, pages 478--483. IEEE, 2007.

\bibitem{burget2008combination}
Burget Lukas, Schwarz Petr, Matejka Pavel, Hannemann Mirko, Rastrow Ariya,
  White Christopher, Khudanpur Sanjeev, Hermansky Hynek, and Cernocky Jan.
\newblock Combination of strongly and weakly constrained recognizers for
  reliable detection of oovs.
\newblock In {\em 2008 IEEE International Conference on Acoustics, Speech and
  Signal Processing}, pages 4081--4084. IEEE, 2008.

\bibitem{kombrink2009posterior}
Kombrink Stefan, Burget Luk{\'a}{\v{s}}, Mat{\v{e}}jka Pavel, Karafi{\'a}t
  Martin, and Hermansky Hynek.
\newblock Posterior-based out of vocabulary word detection in telephone speech.
\newblock In {\em Tenth Annual Conference of the International Speech
  Communication Association}, 2009.

\bibitem{murthy2018effect}
Murthy Savitha, Sitaram Dinkar, and Sitaram Sunayana.
\newblock Effect of tts generated audio on oov detection and word error rate in
  asr for low-resource languages.
\newblock In {\em Interspeech}, pages 1026--1030, 2018.

\bibitem{hori2017multi}
Hori Takaaki, Watanabe Shinji, and Hershey John~R.
\newblock Multi-level language modeling and decoding for open vocabulary
  end-to-end speech recognition.
\newblock In {\em 2017 IEEE Automatic Speech Recognition and Understanding
  Workshop (ASRU)}, pages 287--293. IEEE, 2017.

\bibitem{hannemann2010similarity}
Hannemann Mirko, Kombrink Stefan, Karafi{\'a}t Martin, and Burget
  Luk{\'a}{\v{s}}.
\newblock Similarity scoring for recognizing repeated out-of-vocabulary words.
\newblock In {\em Eleventh Annual Conference of the International Speech
  Communication Association}, 2010.

\bibitem{inbook}
Illina I. and Fohr Dominique.
\newblock {\em RNN Language Model Estimation for Out-of-Vocabulary Words},
  pages 199--211.
\newblock 12 2020.

\bibitem{DBLP:conf/icassp/EgorovaB18}
Egorova Ekaterina and Burget Luk{\'{a}}s.
\newblock Out-of-vocabulary word recovery using fst-based subword unit
  clustering in a hybrid {ASR} system.
\newblock In {\em 2018 {IEEE} International Conference on Acoustics, Speech and
  Signal Processing, {ICASSP} 2018, Calgary, AB, Canada, April 15-20, 2018},
  2018.

\bibitem{DBLP:conf/interspeech/QinSR11}
Qin Long, Sun Ming, and Rudnicky Alexander~I.
\newblock {OOV} detection and recovery using hybrid models with different
  fragments.
\newblock In {\em {INTERSPEECH} 2011, 12th Annual Conference of the
  International Speech Communication Association, Florence, Italy, August
  27-31, 2011}, 2011.

\bibitem{DBLP:journals/ieicet/NaptaliTN12}
Naptali Welly, Tsuchiya Masatoshi, and Nakagawa Seiichi.
\newblock Class-based n-gram language model for new words using
  out-of-vocabulary to in-vocabulary similarity.
\newblock {\em {IEICE} Trans. Inf. Syst.}, 95-D(9):2308--2317, 2012.

\bibitem{DBLP:conf/icnlsp/OrosanuJ15}
Orosanu Luiza and Jouvet Denis.
\newblock Adding new words into a language model using parameters of known
  words with similar behavior.
\newblock In Abbas Mourad and Abdelali Ahmed, editors, {\em 1st International
  Conference on Natural Language and Speech Processing, {ICNLSP} 2015, Algiers,
  Algeria, October 18-19, 2015}, volume 128 of {\em Procedia Computer Science},
  pages 18--24. Elsevier, 2015.

\bibitem{DBLP:conf/icassp/WangZLWDQ21}
Wang Wei, Zhou Zhikai, Lu~Yizhou, Wang Hongji, Du~Chenpeng, and Qian Yanmin.
\newblock Towards data selection on {TTS} data for children's speech
  recognition.
\newblock In {\em {IEEE} International Conference on Acoustics, Speech and
  Signal Processing, {ICASSP} 2021, Toronto, ON, Canada, June 6-11, 2021},
  2021.

\bibitem{DBLP:conf/icassp/ZhengLGW21}
Zheng Xianrui, Liu Yulan, Gunceler Deniz, and Willett Daniel.
\newblock Using synthetic audio to improve the recognition of out-of-vocabulary
  words in end-to-end asr systems.
\newblock In {\em {IEEE} International Conference on Acoustics, Speech and
  Signal Processing, {ICASSP} 2021, Toronto, ON, Canada, June 6-11, 2021},
  2021.

\bibitem{shetty2021exploring}
Shetty Vishwas~M and Umesh S.
\newblock Exploring the use of common label set to improve speech recognition
  of low resource indian languages.
\newblock In {\em ICASSP 2021-2021 IEEE International Conference on Acoustics,
  Speech and Signal Processing (ICASSP)}, pages 7228--7232. IEEE, 2021.

\bibitem{creutz2007morph}
Creutz Mathias, Hirsim{\"a}ki Teemu, Kurimo Mikko, Puurula Antti, Pylkk{\"o}nen
  Janne, Siivola Vesa, Varjokallio Matti, Arisoy Ebru, Sara{\c{c}}lar Murat,
  and Stolcke Andreas.
\newblock Morph-based speech recognition and modeling of out-of-vocabulary
  words across languages.
\newblock {\em ACM Transactions on Speech and Language Processing (TSLP)},
  5(1):1--29, 2007.

\bibitem{lileikyte2018conversational}
Lileikyt{\.e} Rasa, Lamel Lori, Gauvain Jean-Luc, and Gorin Arseniy.
\newblock Conversational telephone speech recognition for lithuanian.
\newblock {\em Computer Speech \& Language}, 49:71--82, 2018.

\bibitem{he2014subword}
He~Yanzhang, Hutchinson Brian, Baumann Peter, Ostendorf Mari, Fosler-Lussier
  Eric, and Pierrehumbert Janet.
\newblock Subword-based modeling for handling oov words inkeyword spotting.
\newblock In {\em 2014 IEEE International Conference on Acoustics, Speech and
  Signal Processing (ICASSP)}, pages 7864--7868. IEEE, 2014.

\bibitem{choueiter2006morpheme}
Choueiter Ghinwa, Povey Daniel, Chen Stanley~F, and Zweig Geoffrey.
\newblock Morpheme-based language modeling for arabic lvcsr.
\newblock In {\em 2006 IEEE International Conference on Acoustics Speech and
  Signal Processing Proceedings}, volume~1, pages I--I. IEEE, 2006.

\bibitem{povey2012generating}
Povey Daniel, Hannemann Mirko, Boulianne Gilles, Burget Luk{\'a}{\v{s}},
  Ghoshal Arnab, Janda Milo{\v{s}}, Karafi{\'a}t Martin, Kombrink Stefan,
  Motl{\'\i}{\v{c}}ek Petr, Qian Yanmin, and others .
\newblock Generating exact lattices in the wfst framework.
\newblock In {\em 2012 IEEE International Conference on Acoustics, Speech and
  Signal Processing (ICASSP)}, pages 4213--4216. IEEE, 2012.

\bibitem{srivastava2018interspeech}
Srivastava Brij Mohan~Lal, Sitaram Sunayana, Mehta Rupesh~Kumar, Mohan
  Krishna~Doss, Matani Pallavi, Satpal Sandeepkumar, Bali Kalika, Srikanth
  Radhakrishnan, and Nayak Niranjan.
\newblock Interspeech 2018 low resource automatic speech recognition challenge
  for indian languages.
\newblock In {\em SLTU}, pages 11--14, 2018.

\bibitem{povey2018semi}
Povey Daniel, Cheng Gaofeng, Wang Yiming, Li~Ke, Xu~Hainan, Yarmohammadi Mahsa,
  and Khudanpur Sanjeev.
\newblock Semi-orthogonal low-rank matrix factorization for deep neural
  networks.
\newblock In {\em Interspeech}, pages 3743--3747, 2018.

\bibitem{povey2011kaldi}
Povey Daniel, Ghoshal Arnab, Boulianne Gilles, Burget Lukas, Glembek Ondrej,
  Goel Nagendra, Hannemann Mirko, Motlicek Petr, Qian Yanmin, Schwarz Petr, and
  others .
\newblock The kaldi speech recognition toolkit.
\newblock In {\em IEEE 2011 workshop on automatic speech recognition and
  understanding}, number CONF. IEEE Signal Processing Society, 2011.

\bibitem{pusateri2019connecting}
Pusateri Ernest, Van~Gysel Christophe, Botros Rami, Badaskar Sameer, Hannemann
  Mirko, Oualil Youssef, and Oparin Ilya.
\newblock Connecting and comparing language model interpolation techniques.
\newblock {\em arXiv preprint arXiv:1908.09738}, 2019.

\bibitem{hsu2007generalized}
Hsu Bo-June.
\newblock Generalized linear interpolation of language models.
\newblock In {\em 2007 IEEE workshop on automatic speech recognition \&
  understanding (ASRU)}, pages 136--140. IEEE, 2007.

\bibitem{chen2020improving}
Chen Zhehuai, Rosenberg Andrew, Zhang Yu, Wang Gary, Ramabhadran Bhuvana, and
  Moreno Pedro~J.
\newblock Improving speech recognition using gan-based speech synthesis and
  contrastive unspoken text selection.
\newblock In {\em INTERSPEECH}, pages 556--560, 2020.

\bibitem{klakow2000selecting}
Klakow Dietrich.
\newblock Selecting articles from the language model training corpus.
\newblock In {\em 2000 IEEE International Conference on Acoustics, Speech, and
  Signal Processing. Proceedings (Cat. No. 00CH37100)}, volume~3, pages
  1695--1698. IEEE, 2000.

\bibitem{itoh2012n}
Itoh Nobuyasu, Sainath Tara~N, Jiang Dan~Ning, Zhou Jie, and Ramabhadran
  Bhuvana.
\newblock N-best entropy based data selection for acoustic modeling.
\newblock In {\em 2012 IEEE International Conference on Acoustics, Speech and
  Signal Processing (ICASSP)}, pages 4133--4136. IEEE, 2012.

\end{thebibliography}

\end{document}